%% file: main.tex
\documentclass{article}
\usepackage{graphicx} 
\usepackage{cite}
\usepackage{amsmath,amssymb,amsfonts}
\usepackage{algorithmic}
\usepackage{graphicx}
\usepackage{textcomp}
\usepackage{chapterbib}  
\usepackage{caption}
\usepackage{subcaption}
\usepackage{array}
\usepackage{multirow}
\usepackage{caption}
\usepackage{subcaption}
\usepackage{booktabs}
\usepackage{multirow}
\usepackage{hyperref}
\usepackage[inkscapelatex=false]{svg}
\usepackage{pdfpages}
\usepackage{geometry}
 \usepackage{float}

\usepackage{authblk}
\makeatletter
\newcommand\footnoteref[1]{\protected@xdef\@thefnmark{\ref{#1}}\@footnotemark}
\makeatother

\title{\textbf{Regional quality estimation for echocardiography using deep learning}}
\author[1]{Gilles Van De Vyver}
\author[1]{Svein-Erik Måsøy}
\author[1,2]{Håvard Dalen}
\author[1,2]{Bjørnar Leangen Grenne}
\author[1,2]{Espen Holte}
\author[1]{Sindre Hellum Olaisen}
\author[1]{John Nyberg}
\author[1,3]{Andreas Østvik}
\author[1]{Lasse Løvstakken}
\author[1.3]{Erik Smistad}
\affil[1]{Norwegian University of Science and Technology, Trondheim, Norway} 
\affil[2]{St. Olavs hospital, Trondheim, Norway} 
\affil[3]{Health Research, SINTEF, Trondheim, Norway} 
\date{}

\begin{document}

\maketitle
\begin{abstract}
Automatic estimation of cardiac ultrasound image quality can be beneficial for guiding operators and ensuring the accuracy of clinical measurements. Previous work often fails to distinguish the view correctness of the echocardiogram from the image quality. Additionally, previous studies only provide a global image quality value, which limits their practical utility. In this work, we developed and compared three methods to estimate image quality: 1) classic pixel-based metrics like the generalized contrast-to-noise ratio (gCNR) on myocardial segments as region of interest and left ventricle lumen as background, obtained using a U-Net segmentation 2) local image coherence derived from a U-Net model that predicts coherence from B-Mode images 3) a deep convolutional network that predicts the quality of each region directly in an end-to-end fashion. We evaluate each method against manual regional image quality annotations by three experienced cardiologists. The results indicate poor performance of the gCNR metric, with Spearman correlation to the annotations of $\rho=0.24$. The end-to-end learning model obtains the best result, $\rho=0.69$, comparable to the inter-observer correlation, $\rho=0.63$. Finally, the coherence-based method, with $\rho=0.58$, outperformed the classical metrics and is more generic than the end-to-end approach. The image quality prediction tool is available as an open source Python library at \url{https://github.com/GillesVanDeVyver/arqee}.
\begin{center}
    \textbf{Keywords}:
\end{center}
Cardiac segmentation, Ultrasound, image quality, Coherence, Signal-To-Noise Ratio

\end{abstract}

\input{Sections/Introduction}

\input{Sections/Datasets}

\input{Sections/Methods}

\input{Sections/Experimental_setup_and_Results}

\input{Sections/Discussion}

\input{Sections/Conclusion}

\bibliographystyle{IEEEtran}

\bibliography{bibliography.bib}

\input{Sections/appendices}

\end{document}

%% file: Sections/Introduction.tex
\section{Introduction}

Image quality is one of the main challenges in ultrasound imaging and can differ significantly between patients and imaging equipment. In echocardiography, many factors influence image quality such as the ultrasound scanner, the patient, and the probe. Several quantitative measurements using the images are performed. However, this requires image quality good enough for the given measurement. Different measurements have different image quality requirements, for instance, left ventricular (LV) volume, ejection fraction (EF), and strain measurements require good image quality in the entire myocardium. On the other hand, mitral annular plane systolic excursion (MAPSE) only requires good image quality in the annulus. Good image quality should generally provide measurement values with low uncertainty. Estimating image quality can be useful in the following ways:
\begin{itemize}
    \item To guide operators to achieve as good image quality as possible while scanning.
    \item To automatically select the best images, recordings, and the best cardiac cycles to use for a given measurement.
    \item As quality assurance, e.g. to warn the user when an image is not good enough for a measurement, and to automatically approve/disapprove individual myocardial segments based on image quality.
    \item In data mining projects, to exclude cases with insufficient quality for reliable measurements.
\end{itemize}

We distinguish between two types of quality of ultrasound images: view quality/correctness and image quality. 
In this work, we will focus on image quality specifically. For view correctness, previous work has demonstrated that 3D ultrasound can serve as training data to automatically identify the transducer rotation and tilt in relation to the desired standard view and can guide the user to the correct position \cite{pasdeloup2023real, droste2020automatic, ostvik2019real}. 
For image quality, the classic ultrasound signal-processing metrics are the contrast ratio (CR) \cite{smith1985frequency}, contrast-to-noise ratio (CNR) \cite{patterson1984improvement}, and generalized CNR (gCNR) \cite{rodriguez2019generalized}. These three metrics need a region of interest (ROI) and a background region to compare against. More recently, global image coherence (GIC) \cite{rindal2023very} has been proposed as a general quality metric that does not require the selection of these two regions. The image coherence measures how well the signals of the transducer elements align after delay compensation, with more alignment corresponding to clearer and sharper images. From the above mentioned methods, only the GIC can be used directly and automatically for measuring image quality separately as it does not require selecting an ROI and noise region. However, this approach requires channel data, which is not readily available in practice and does not give regional metrics.  \newline

Several automatic methods for measuring ultrasound image quality have been published applicable to cardiac imaging. Abdi et al. \cite{abdi2017quality} used a recurrent neural network to predict the global quality of cardiac cine loops. The criteria for quality assessment take both image quality and view correctness into account. In subsequent studies \cite{luong2021automated, van2018quantitative}, they used an architecture that performs both view classification and global quality prediction simultaneously. The image quality metric is a global criterion based on the manual judgment of the clarity of the blood-tissue interface. Labs et al. \cite{labs2023automated} used a multi-stream neural network architecture where each stream takes in a sequence of frames and predicts a specific quality criterion. The criteria are global and take both view correctness and image quality into account. Karamalis et al. \cite{karamalis2012ultrasound} detect attenuated shadow regions with random walks resulting in a pixel-level confidence map. Unlike the other methods above, this method is not based on deep learning. It provides a local, pixel-level metric, but it only measures the visibility of regions and not the quality of their content. \newline

All of the automatic methods mentioned above have the limitation that they only provide a global image quality evaluation and/or do not assess the image quality separate from the view correctness.  \newline

%% file: Sections/Datasets.tex
\section{Methods}
In this work, we developed and compared three fully automatic methods to asses regional image quality in cardiac ultrasound separate from the view correctness:
\begin{itemize}
    \item Classical ultrasound image quality metrics, such as contrast ratio and contrast to noise ratio, applied in cardiac regions automatically extracted using deep-learning segmentation.
    \item Deep-learning predicted ultrasound coherence, which is a measure of how coherent a signal is received by the transducer elements, together with deep-learning segmentation.
    \item End-to-end prediction of regional image quality.
\end{itemize}

The rest of this section first presents the datasets used to develop these methods, and then presents each of the three methods.

\subsection{Datasets}

\subsubsection{VLCD}

The \textbf{Very Large Cardiac Channel Data Database (VLCD)} consists of channel data from 33280 frames from 538 recordings of 106 study participants \cite{rindal2023very}. It contains parasternal short axis (PSAX), parasternal long axis (PLAX), apical long axis (ALAX), apical two-chamber (A2C), and apical four-chamber (A4C) views. We split the VLCD dataset on the study participant level into train, validation, and test sets, 70\%, 15\%, and 15\% respectively. 

\subsubsection{HUNT4}

The Nord-Trøndelag Health Study dataset (HUNT4Echo) is a clinical dataset including among others PSAX, PLAX, ALAX, A2C, and A4C views. Each recording contains 3 cardiac cycles. We use two subsets of the HUNT4Echo dataset.
\begin{itemize}
    \item \textbf{Segmentation annotation dataset} A fraction of 311 study participant exams, the segmentation annotation set \cite{olaisen2023automatic}, contains single frame segmentation annotations in both ED and ES as pixel-wise labels of the left ventricle (LV), left atrium (LA), and myocardium (MYO) in ALAX, A2C, and A4C views.
    \item \textbf{Regional image quality dataset} For this work, we created an additional dataset of image quality labels. The local image quality labels are manual annotations that asses the image quality of the cardiac regions of interest on a subset of the HUNT4 dataset in ALAX, A2C, and A4C views. Section \ref{section: Regional image quality annotation on HUNT4} describes the annotation process in more detail.
\end{itemize}

%% file: Sections/Methods.tex
\subsection{Regional image quality annotation on HUNT4} \label{subsubsection: Regional image quality annotation} \label{section: Regional image quality annotation on HUNT4}
An annotation tool was developed specifically for this project using the open-source Annotation Web software\footnote{\url{https://github.com/smistad/annotationweb}} \cite{smistad2021annotation}.
The tool was made to enable clinicians to annotate regional image quality as efficiently and accurately as possible. The tool is freely available and can be adapted to other image quality projects. The image quality annotations were performed by three experienced cardiologists using the following protocol:
\begin{enumerate}
\item Annotate the end-diastole (ED) and end-systole (ES) frame of each recording, and optionally other frames if the image quality changes significantly during the recording.
\item If the majority of the cardiac regions of interest is out-of-sector, label it as out-of-sector. Otherwise, label the part of the region that is inside the sector according to the definitions in Table \ref{table: def_labels}. We ignore the out-of-sector regions in the remainder of this work.
\end {enumerate}
For the first round of annotations, each of the three clinicians annotated the same 10 frames from 5 recordings of 2 study participants. We used this dataset to calculate the inter-observer variability. For the second round of annotations, the three clinicians collectively annotated 458 frames from 158 recordings of 65 study participants. The annotations from the second round form the \textbf{regional image quality dataset}. This dataset was split randomly at the study participant level into train, validation, and test sets, allocating 70\%, 15\%, and 15\% of the data to each set respectively.

\begin{table}
\scriptsize
  \centering
  \caption{Definitions of image quality labels for annotating cardiac regions of interest. We only look at the signal quality of the region that is inside the ultrasound sector. If more than 50\% is outside the sector, the region is treated as out-of-sector and excluded from analysis. }
  \begin{tabular}{m{40pt}|m{360pt}|m{40pt}}
    \toprule
    image quality label &  Definition & Quality score\\
\midrule
Not visible & Less than 50\% of what is inside the sector is visible. & 1 \\ & & \\
Poor &  Low signal-to-noise ratio and/or smearing/low resolution, but can still see at least 50\% of the myocardium/annulus. Diffuse border between annulus and cavity. & 2 \\ & &
\\
Ok & Medium signal-to-noise ratio and/or some smearing/lower resolution, can see at least 70\% of the myocardium/annulus. & 3 \\ & &
\\
Good & Good signal-to-noise ratio and clear border in at least 80\% of the myocardium/annulus. Very little smearing/good resolution. Clear border between annulus and cavity & 4 \\ & &
\\ 
Excellent & Within the best 10\% of what you have seen on this system. & 5 \\ & &
\\
\bottomrule
  \end{tabular}
      \label{table: def_labels}
\end{table}

\subsection{Regional image quality estimation}

\subsubsection{Classical ultrasound image quality metrics}
For the classical image quality metrics, deep-learning segmentation is used to extract the annulus regions and each of the myocardial segments as regions of interest and the LV as the background region. Appendix \ref{appendix: Extraction of cardiac regions of interest} gives more details about the procedure for dividing the segmentation into regions. The four classical ultrasound image quality metrics below were tested.
We apply histogram matching \cite{bottenus2020histogram, gonzalez2009digital} to a Gaussian distribution ($\mu=127 , \sigma=32$) for the B-Mode grayscale images before calculating pixel-based quality metrics.

\begin{enumerate}

  \item[\textbullet] \textbf{Pixel intensity} is the average pixel intensity value in each region.

  \item[\textbullet] \textbf{Contrast Ratio (\textit{CR})} \cite{smith1985frequency} is defined as 
  \[ CR=\frac{\mu_{\text{segment}}}{\mu_{\text{LV}}}\]
  where $\mu_{\text{segment}}$ is the average intensity in each region and $\mu_{\text{LV}}$ is the average intensity inside the LV lumen.
  
  \item[\textbullet] \textbf{Contrast to Noise Ratio (\textit{CNR}) \cite{patterson1984improvement}} is defined as 
    \[ CNR=\frac{\mu_{\text{segment}}-\mu_{\text{LV}}}{\sqrt{\sigma_{\text{segment}}^2+\sigma_{\text{LV}}^2}}\]
    where $\sigma_{\text{segment}}$ is the standard deviation in each region and $\sigma_{\text{LV}}$ is the standard deviation inside the LV lumen.

  \item[\textbullet] \textbf{Generalized CNR (\textit{gCNR})} \cite{rodriguez2019generalized} is defined as the maximum performance that can be expected from a hypothetical pixel classifier based on intensity using a set of optimal thresholds. It is calculated as 
\[gCNR= 1-\frac{1}{2}\sum_{i=0}^{MAX_{i}} \min\{p_{\text{segment}}(i), p_{\text{LV}}(i)\} \]
where $p_{\text{segment}}(x)$ is the probability density function of the pixel intensities inside the region the \textit{gCNR} is calculated for, $p_{\text{LV}}(x)$ the probability density function of the pixel intensities inside the LV lumen, and $MAX_{i}$ the maximum possible pixel intensity. 
~Fig. \ref{fig: otsu_thresholding_example_density} shows an example of the probability density functions for one of the regions as ROI and the LV lumen as background.

\begin{figure}
     \centering
     \begin{subfigure}[b]{0.5\linewidth}
         \centering \includegraphics[trim={0cm 0cm 0cm 0cm},clip,width=1\linewidth]{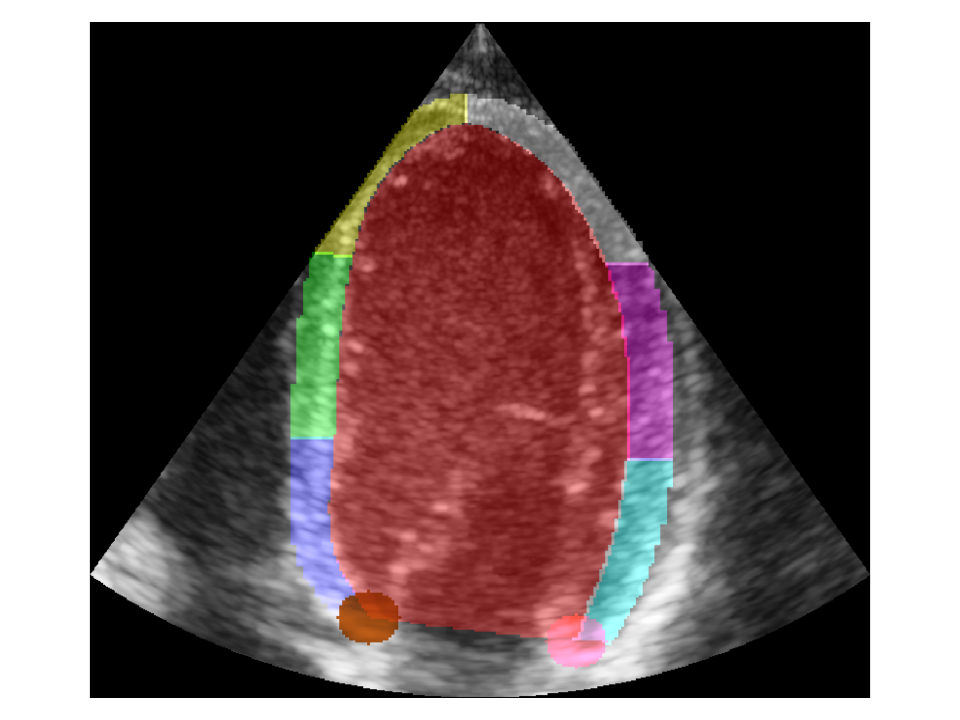}
         \caption{ \centering Segmentation}
         \label{fig: otsu_thresholding_example_density_a}
     \end{subfigure}
          \hspace*{-0.75cm}
     \begin{subfigure}[b]{0.5\linewidth}
         \centering \includegraphics[trim={0cm 0.5cm 0cm 0.cm},clip,width=1\linewidth]{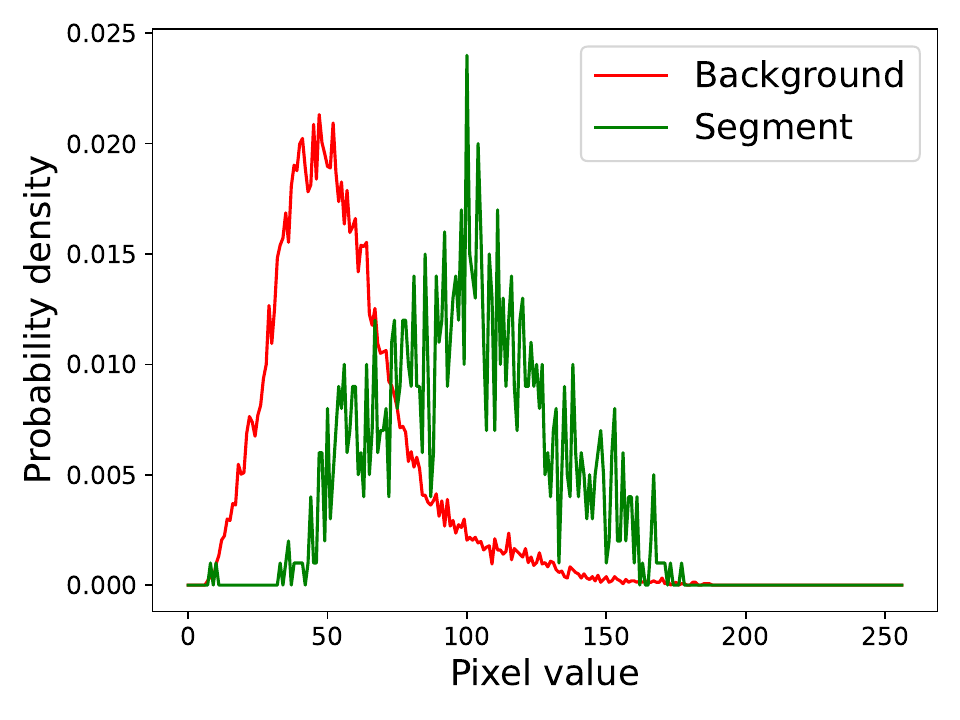}
         \caption{\centering Probability density functions}
         \label{fig: otsu_thresholding_example_density_b}
     \end{subfigure}
        \caption{Calculating gCNR for a region of the myocardium. The left side of the figure shows the segmentation with the MYO and annulus points divided into regions. The right side shows the probability density functions of the segment and background pixels used to calculate the gCNR. In this example, we use the mid region on the left side as ROI and the full LV lumen as background. These correspond to the green and red masks respectively in the left part of the figure.}
        \label{fig: otsu_thresholding_example_density}
    \hspace{1cm}
\end{figure}

\end{enumerate}

\subsubsection{Local, deep-learning predicted image coherence as quality metric} \label{section: Coherence prediction from B-mode}

We use the VLCD dataset to calculate the coherence factor \cite{rigby1999method} for each pixel in the ultrasound image. This factor is the ratio between the amplitude of the sum of the received signals to the sum of the amplitudes of those signals, 
\[ CF = \frac{\sum_{n=1}^{N}S_{i}}{\sum_{n=1}^{N}|S_{i}|} \]
where $S_{i}$ is the delayed signal for the \textit{i}-th transducer element. This is equivalent to taking the coherent sum of the signal and dividing it by the incoherent sum of each signal. Thus, the coherence factor measures of how well the complex signals of all transducer elements align. The remainder of the signal-processing chain is the native processing of the GE HealthCare Vivid E95 system\footnote{\label{note: signal processing} Gundersen et al. \cite{gundersen2024hardware} describe this signal-processing chain in more detail.} but without the log compression. The result is a \textbf{coherence image} with the same dimension as the B-Mode image.  The final preprocessing step applies gamma normalization with $\gamma=0.5$ on the coherence images,
\[t_{i,normalized}=t_{i}^{\gamma}\]
where $t_{i}$ are the pixels of the target coherence image. The corresponding B-mode images are generated from the channel data using the same, native signal-processing pipeline. \newline

The HUNT4 dataset, like most ultrasound datasets, does not include channel data. Therefore, VLCD was used to train an image-to-image network that takes as input the grayscale B-mode image and predicts the coherence image, which is then used to calculate local image coherence. We use a lightweight U-Net architecture inspired by the U-Net 1 architecture in \cite{leclerc2019deep}, with characteristics listed in Table \ref{table: characteristics coh u-net}. As coherence is related to image quality, we only apply augmentations that do not influence the quality of the image. Furthermore, the coherence should be invariant to different gain settings, so we additionally augment with brightness adjustments on the B-mode image while keeping the target coherence image unchanged. During training and validation of the coherence prediction model, we sample a random frame from each recording in the train and validation set respectively during each epoch. During testing, we use all frames in the test set. The local image coherence quality metric of a region is the average pixel value of all pixels corresponding to the region in the coherence image. This is the same as the pixel intensity metric above but applied to the coherence image instead of the B-Mode image.

\begin{table}
\scriptsize
  \centering
  \caption{Characteristics of the coherence prediction network. The general architecture is a U-Net. The "number of channels" row indicates the number of channels at the first, bottom, and last convolution of the U-Net respectively.}
  \begin{tabular}{m{100pt}|m{250pt}}
    \toprule
    Input size & $256\times256$ \\
    Number of channels & 16 $\downarrow$ 128 $\uparrow$ 16 \\ 
    Number of output channels & 1 \\
    Lowest resolution & $4\times4$ \\ 
    Upsampling scheme & $2\times2$ repeats \\ 
    Normalization scheme & BatchNorm \\ 
    Batch Size & 32 \\
    Optimizer & Adam \\
    Initial learning Rate & 1e-2 \\ 
    Scheduler & None \\ 
    Loss function & Negative structural similarity index measure (SSIM) \cite{wang2004image} \\ 
    Inter-layer Activation & Gaussian error linear unit (GELU) \cite{hendrycks2016gaussian}\\ 
    Final layer Activation & Sigmoid \\ 
    Epochs & 500 \\
Augmentations & Rotations ($-45^{\circ} \leq \text{angle} \leq  45^{\circ}$), horizontal mirroring, gamma correction ($0.75\leq \gamma \leq 1.25$, only on B-mode), scaling ($0.75 \leq \text{magnification} \leq 1.25$), contrast/brightness adjustments ($0.2\leq \text{gain} \leq 2$, $0\leq \text{bias} \leq 128$, only on B-mode). Each augmentation is applied individually with $0.5$ probability. \\
        \bottomrule
  \end{tabular}
      \label{table: characteristics coh u-net}
\end{table}

\subsubsection{End-to-end deep-learning quality prediction} \label{section: End to end quality prediction}

The end-to-end learning approach trains a convolutional neural network on the regional image quality dataset to predict the quality of each region directly. We treat the problem as a regression task where the model predicts a score for each segment. Table \ref{table: def_labels} shows the correspondence between quality scores and annotation labels. As architecture, we use a MobileNetV2 \cite{sandler2018mobilenetv2} that predicts the image quality labels of all regions simultaneously. Appendix \ref{appendix: Ablation study end-to-end learning model} describes the ablation study conducted to justify this setup. \newline

%% file: Sections/Experimental_setup_and_Results.tex
\section{Experimental setup}

\subsection{Evaluation of coherence prediction from B-mode}

We use the structural similarity index (SSIM) \cite{wang2004image}, peak signal-to-noise ratio (PSNR)\footnote{\label{note: max_pixel_err}The maximum pixel value for coherence images is 1.}, and relative pixel error (RPE) to evaluate the coherence image prediction network. We define the RPE as 
\[ RPE = \frac{|t_{i}-p_{i}|}{max(t_{i},\epsilon)} \]
where $t_{i}$ are the pixel values of the target coherence image, $p_{i}$ the pixel values of the predicted coherence image, and $\epsilon=1e-4$. Section \ref{subsection: Results of coherence prediction from B-mode} contains the results of this experiment.

\subsection{Evaluation of quality metrics}
\label{subsection: Evaluation of quality metrics}

The correlation and accuracy of each quality metric were measured by comparing them to the expert annotations on the test set of the regional image quality dataset. For the classic image quality methods and regional coherence method, linear regression models were used to map the quality metric values to image quality labels. The train and validation set were used together to fit the linear regression model and evaluate it on the test set. Section \ref{subsection: Results of quality metrics} contains the results of this experiment.

\subsection{Comparison to inter-observer variability}

We compare the end-to-end, local image coherence, and gCNR-based model to the inter-observer variability on the data obtained in the first round of annotation, explained in subsection \ref{subsubsection: Regional image quality annotation}. For the local image coherence and gCNR-based models, the same linear model as subsection \ref{subsection: Evaluation of quality metrics} was used to map the metrics to quality labels. The inter-observer variability is calculated from the aggregate of the three unique pairwise score errors between each of the three annotators:
\[
e_{\text{inter-observer}} = e_{12} \cup e_{23} \cup e_{13}
\]
where $e_{ij}$ is the score difference between operator i and j. The error metrics of the automatic methods are calculated from the aggregate of the pairwise score errors between the output of the method and each of the three annotators:
\[
e_{M} = e_{1M} \cup e_{2M} \cup e_{3M}
\]
where $e_{iM}$ is the score difference between operator \textit{i} and method \textit{M}. 
Section \ref{subsection: Results of comparison to inter-observer variability} contains the results of this experiment.

\subsection{Relation to variability in clinical measurements}

This experiment evaluates whether there is a relation between the predicted quality and the agreement between different methods for clinical measurements. The hypothesis is that with lower image quality the variability, and thus the uncertainty, of the measurements between methods and between experts increases. More specifically,
this analysis compares peak global longitudinal strain (GLS) and ejection fraction (EF) measurements obtained either fully automatically with AI tools or manually by using GE HealthCare EchoPAC software on HUNT4 \cite{nyberg2023echocardiographic, olaisen2023automatic}. For AI estimation of GLS and EF, the deep-learning methods proposed by Østvik et al. \cite{ostvik2021myocardial} and
Smistad et al. \cite{9037081} were used respectively. The study participants in HUNT4 used for model development were excluded from the analysis. For GLS, the predicted quality is the average quality of all segments over the full recording.
For EF, the predicted quality is the average quality of all segments in the end-diastole (ED) and end-systole (ES) frames of all cycles in the recording. Section \ref{subsection: Results of relation to variability in clinical measurements} contains the results of this experiment.

\section{Results}

\subsection{Results of coherence prediction from B-mode} \label{subsection: Results of coherence prediction from B-mode}

\begin{figure}
\centering
  \centering
  \includegraphics[trim={0cm 0.2cm 0cm 0 cm}, clip,width = \linewidth]{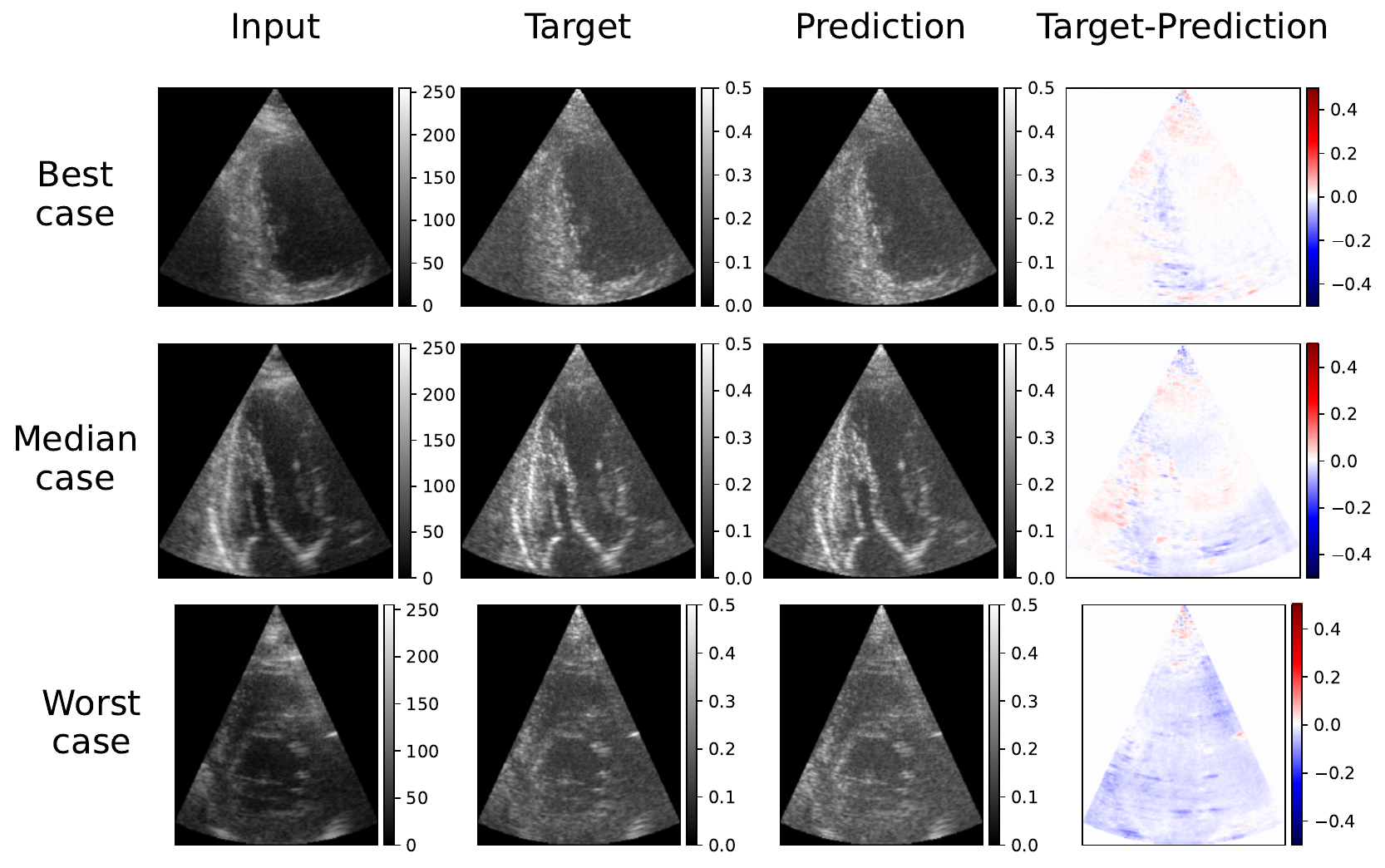}
  \caption{Best, median, and worst case of image-to-image coherence prediction task 
  with relative pixel errors of 4.0, 5.4, and 9.1\% respectively.  The first column shows the input B-mode image. The second column shows the coherence image as predicted by the image-to-image network. The third column shows the ground truth coherence image as calculated from the channel data. Finally, the rightmost column shows the color-coded difference of the target minus the predicted image. 
  }
  \label{fig: best_median_worst}
\end{figure}

\begin{figure}
\centering
  \centering
  \includegraphics[trim={3cm 1cm 4cm 0cm}, clip, width = 0.85\linewidth]{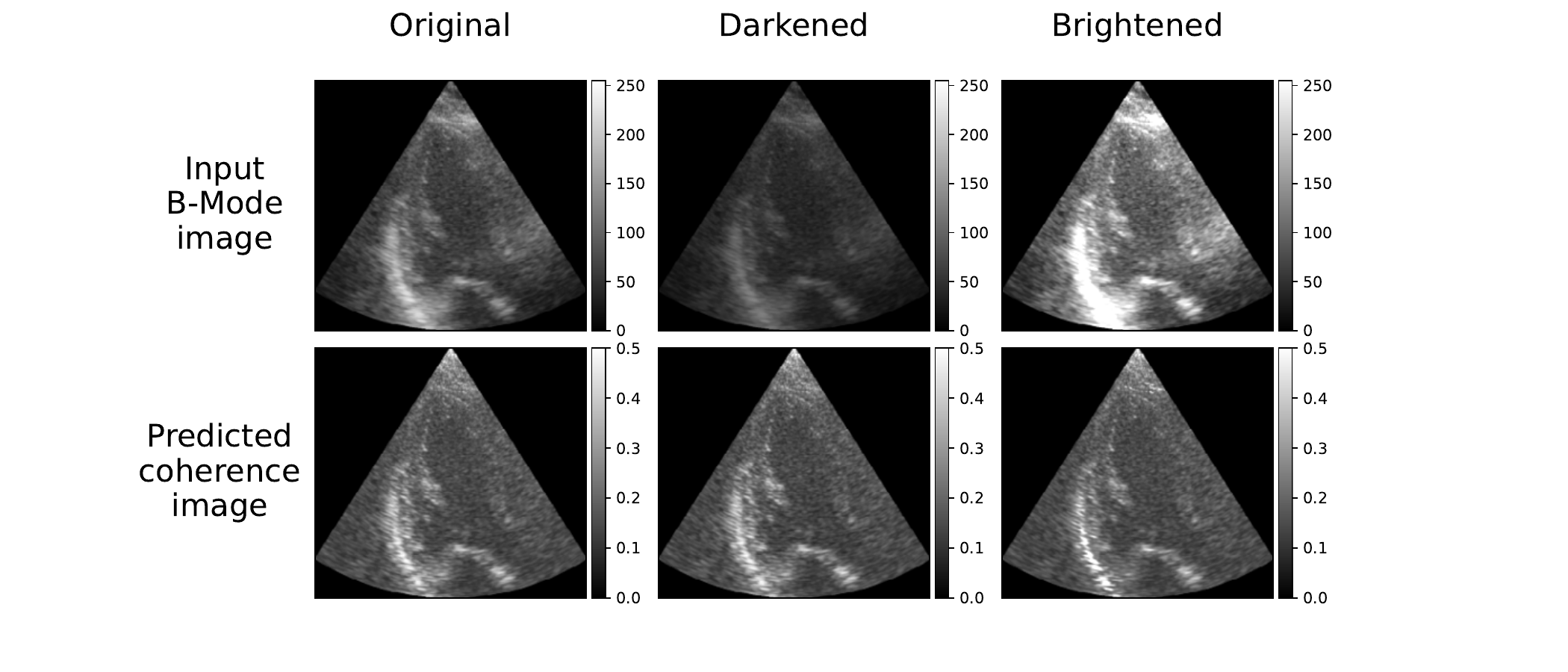}
  \caption{Effect of brightness on coherence prediction. The first row shows a B-Mode image from the regional image quality dataset, brightened and darkened with gamma correction ($\gamma=$ 0.9 and 1.1). The second row shows the predicted coherence images generated by giving the corresponding input from the first row to the network. The predicted coherence is unaffected by the adjustments in brightness, apart from the saturation effect in the brightened image reducing the information in the input, as can be seen in the basal part of the inferolateral wall.}
  \label{fig: coherence_brightness_plot}
\end{figure}

Table \ref{table: results_coh_pred} summarizes the average metric values on the test set. Fig. \ref{fig: best_median_worst} shows an example of the best, median, and worst-case predictions according to the relative pixel error. Fig. \ref{fig: coherence_brightness_plot} illustrates how the predicted coherence images are almost independent of the brightness/gain and contrast/dynamic range of the input B-Mode images. The main finding is that the difference between the estimated and ground truth coherence images is small and thus the predicted coherence images can be used to obtain coherence-based quality metrics for B-mode for which the channel data is not available.

\begin{table}
\scriptsize
  \centering
  \caption{Average metric values of the coherence prediction network on the test set of the VLCD.}
  \begin{tabular}{m{95pt}|m{75pt}}
  \toprule
    SSIM \cite{wang2004image}& $0.994\pm0.003$\\
    Relative pixel error (RPE) & $5.6\%\pm0.89\%$\\
    PSNR\footnoteref{note: max_pixel_err}. & $35.99\pm2.00$ dB\\
    \bottomrule
  \end{tabular}
      \label{table: results_coh_pred}
\end{table}

\subsection{Results of quality metrics} \label{subsection: Results of quality metrics}

\begin{figure}
     \centering
     \begin{subfigure}[b]{0.49\linewidth}
         \centering \includegraphics[clip, width=1\linewidth]{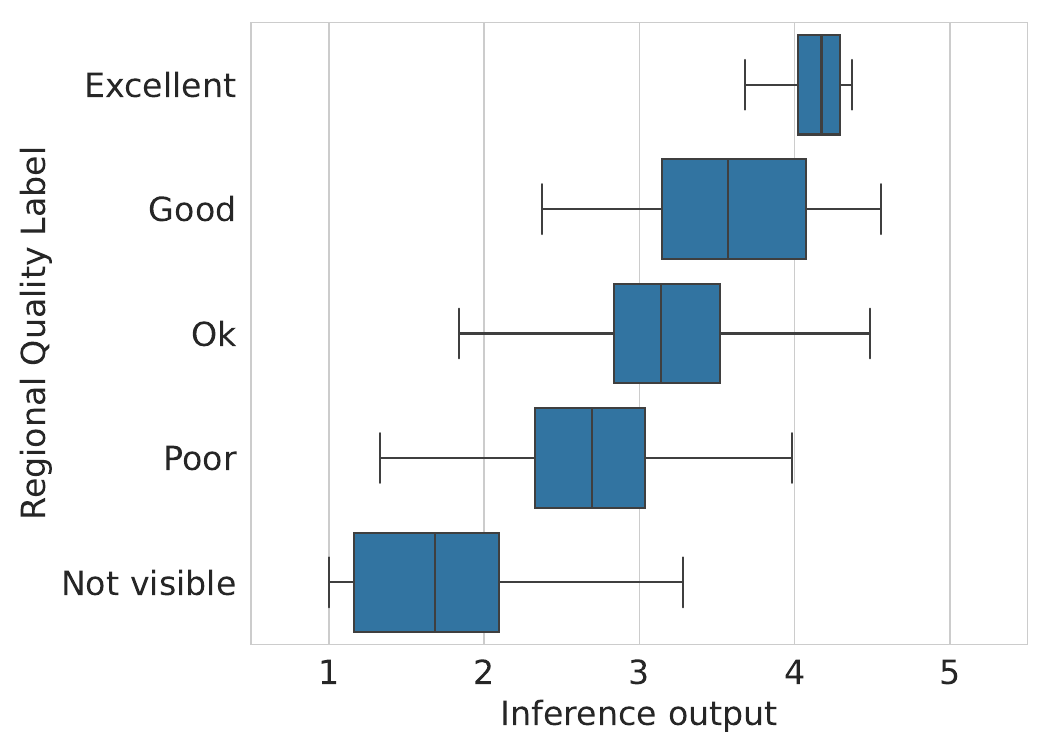}
         \caption{End-to-end model}
         \label{fig: box_plots_end_to_end}
     \end{subfigure}
     \begin{subfigure}[b]{0.49\linewidth}
         \centering \includegraphics[clip,width=1\linewidth]{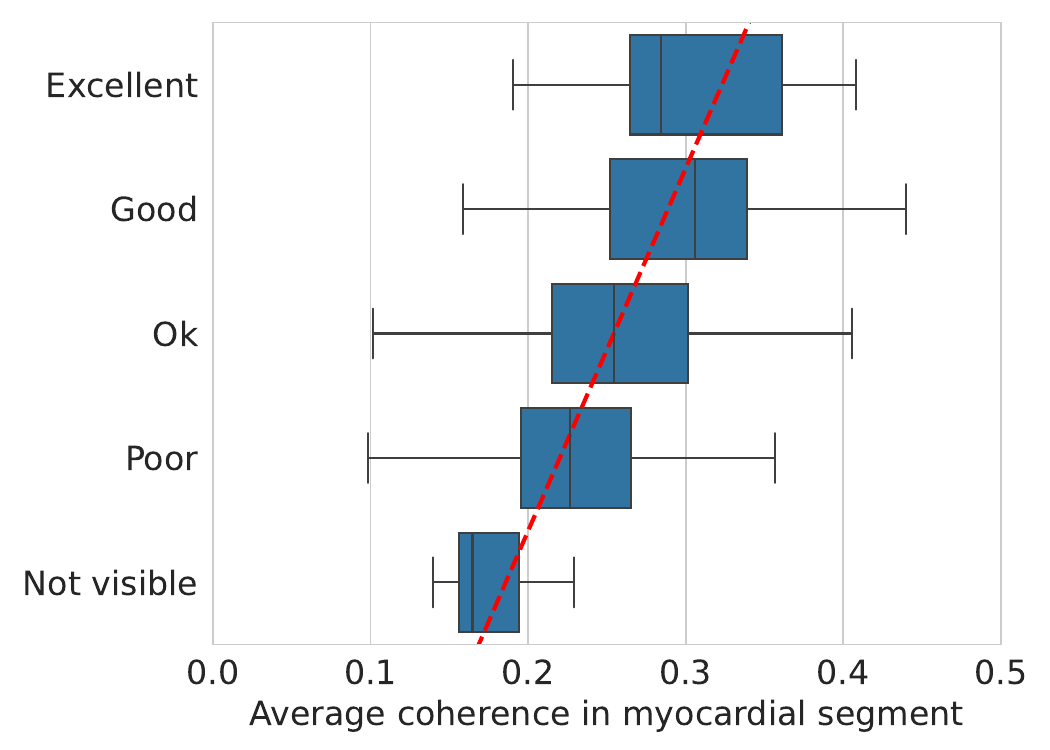}
         \caption{Local image coherence}
         \label{fig: box_plots_coh}
     \end{subfigure}
     \begin{subfigure}[b]{0.49\linewidth}
        \centering \includegraphics[trim={0cm 0cm 0.4cm 0cm},clip,width=1\linewidth]{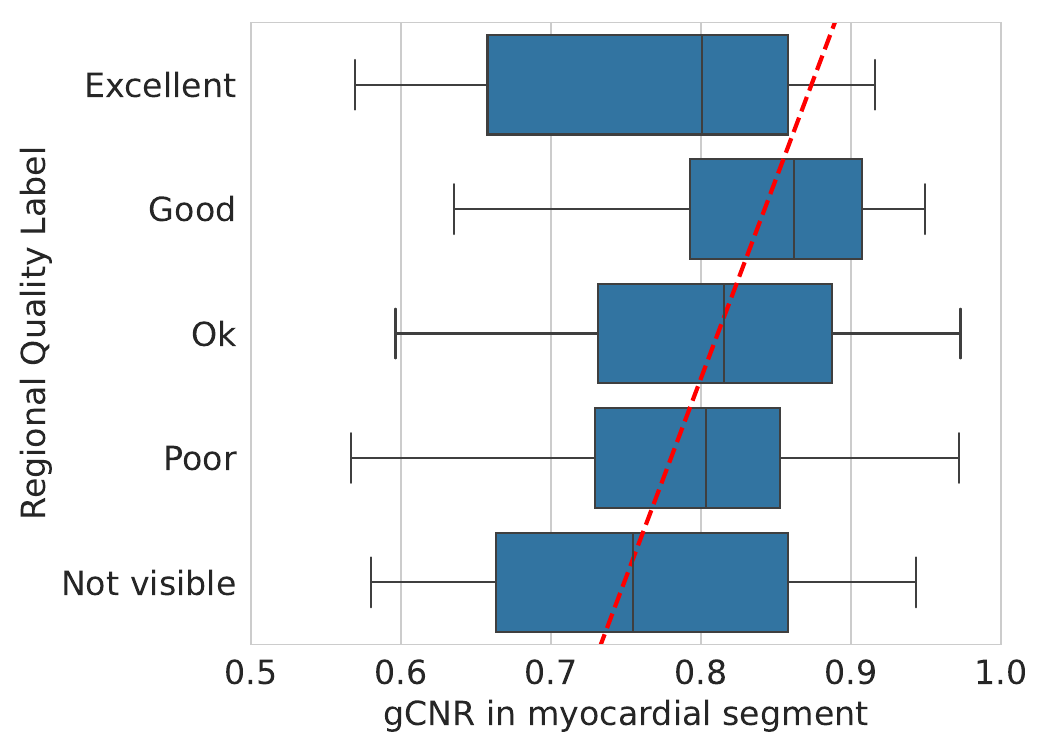}
         \caption{gCNR}
         \label{fig: box_plots_gcnr}
     \end{subfigure}
     \begin{subfigure}[b]{0.49\linewidth}
        \centering \includegraphics[clip,width=1\linewidth]{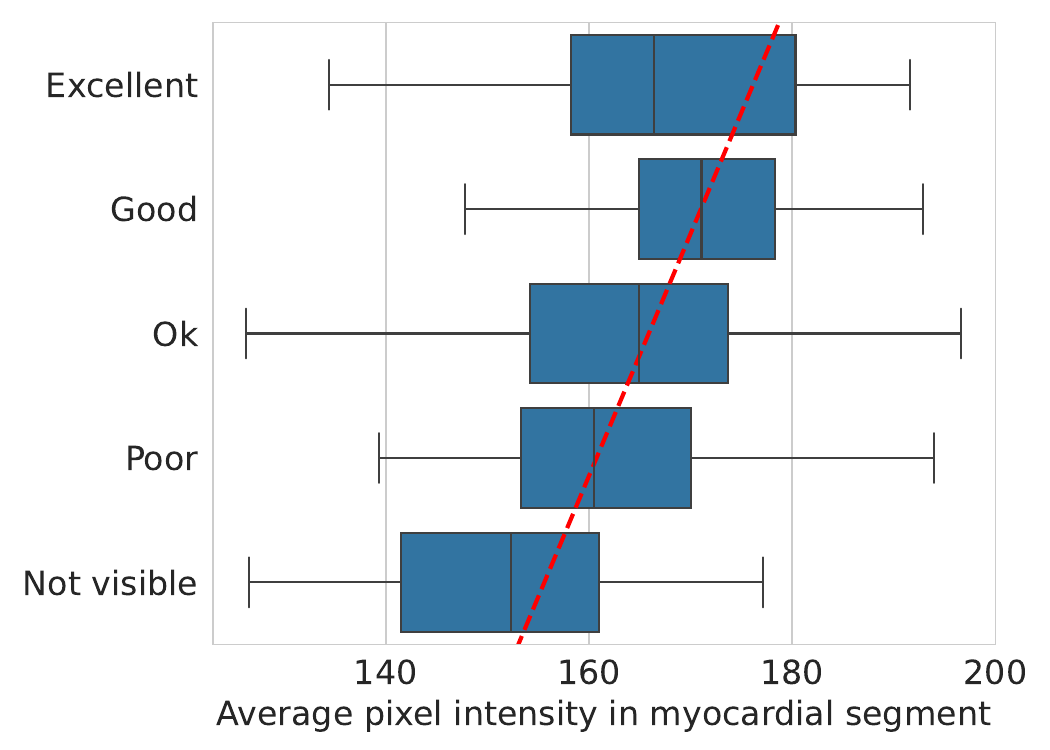}
         \caption{Pixel intensity}
         \label{fig: box_plots_intensity}
     \end{subfigure}  
    \caption{Box plots of quality metrics versus regional quality labels on the test set of the regional image quality dataset. The predictions of the end-to-end model have the strongest correlation to the quality labels. The dotted line represents the linear regression model that maps the quality metrics to quality labels, as described in section \ref{subsection: Evaluation of quality metrics}. The inference output of the end-to-end model can be used directly without additional linear model.}
    \label{fig: box_plots}
\end{figure}

\begin{figure}
\centering
\includegraphics[trim={1cm 5cm 0.8cm 1cm}, clip, width=0.8\linewidth]{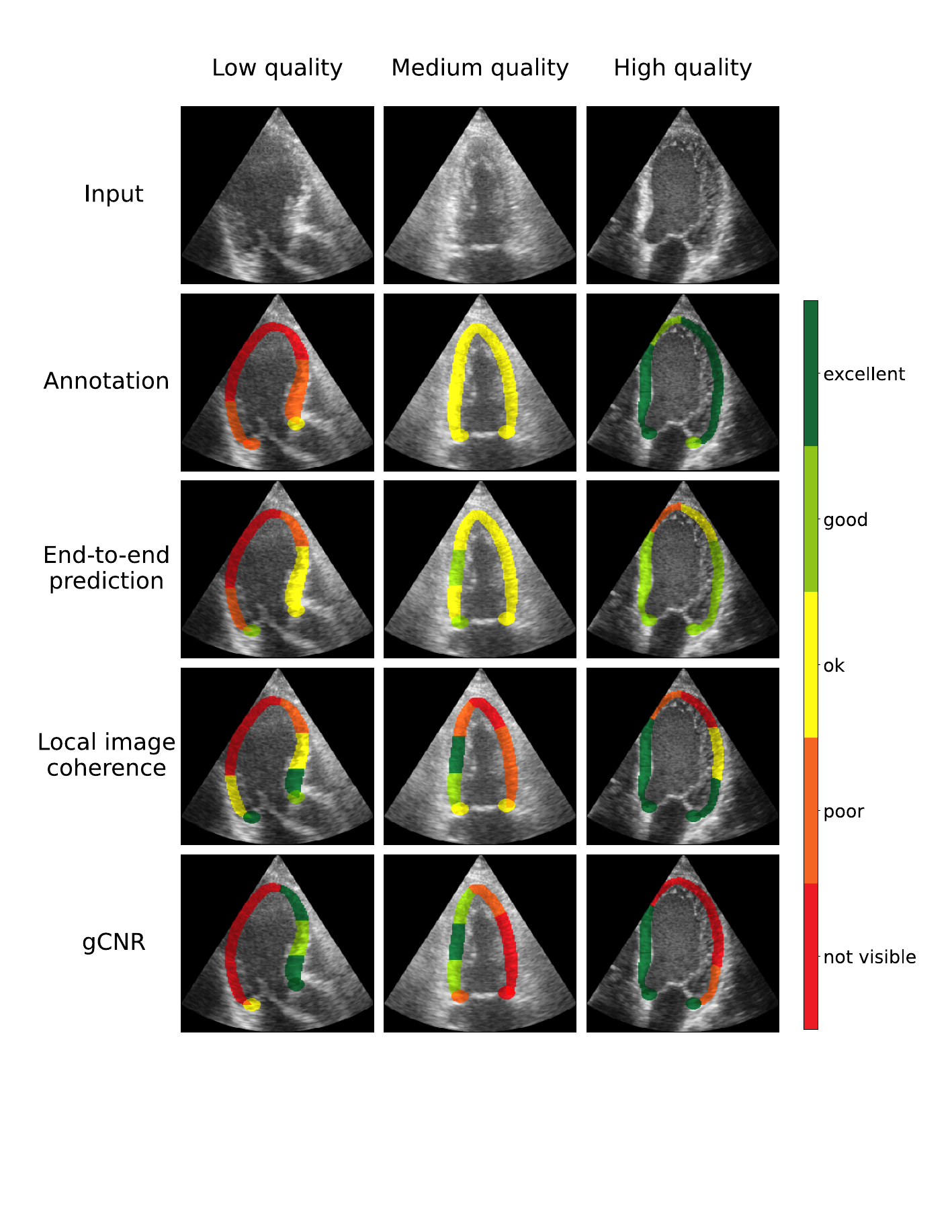}
\caption{Example cases of annotations and automatically predicted regional quality from the test set. The visualization uses the regional quality metrics to color-code the output of divided segmentation output. The end-to-end model predicts the regional qualities directly from the B-mode without using the segmentation output. The local image coherence metric uses the segmentation output to select ROI. The gCNR metric uses the segmentation output to select ROI and background region. The background region, i.e., the LV lumen, is not shown in the image.}
\label{fig: example_quality_cases}
\end{figure}

Table \ref{table: quality_metrics_results} summarizes the results of the evaluation of the quality metrics. Fig. \ref{fig: box_plots} shows box plots of the quality metrics per image quality label for the end-to-end, coherence, gCNR, and intensity models. Fig. \ref{fig: example_quality_cases} shows examples of B-mode images with varying quality together with labels from the annotators and automatic quality metrics. The main finding is that the end-to-end model performs the best, followed by the local image coherence metric. The classical ultrasound image quality metrics perform poorly.

\begin{table}
\scriptsize
  \centering
  \caption{Comparison of quality metrics on the test set of the regional image quality dataset, i.e. the second round of annotations. Accuracy is defined as the ratio of agreement between the automatic measurement, rounded to the nearest quality category, and the annotation on the segment level.}
  \begin{tabular}{m{100pt}|m{100pt}|m{100pt}|m{100pt}}
  \toprule
    Quality metric & Spearman correlation & Mean absolute error (MAE) & Accuracy \\
    \midrule
    Pixel intensity & 0.30 & 1.31$\pm$0.91  & 20\%  \\
    CR & 0.28 & 1.32$\pm$0.92  & 22\% \\
    CNR & 0.25 & 1.38$\pm$0.96 0. & 20\% \\
    gCNR & 0.20 & 1.50$\pm$0.96 & 15\%  \\
    Local image coherence & 0.48 & 1.07$\pm$0.80  & 29\%  \\
    End-to-end learning & \textbf{0.65} & \textbf{0.59$\pm$0.45} & \textbf{48\% } \\
    \bottomrule
  \end{tabular}
      \label{table: quality_metrics_results}
\end{table}

\subsection{Results of comparison to inter-observer variability} \label{subsection: Results of comparison to inter-observer variability}

\begin{figure}
\centering
  \centering
  \includegraphics[width = 0.55\linewidth]{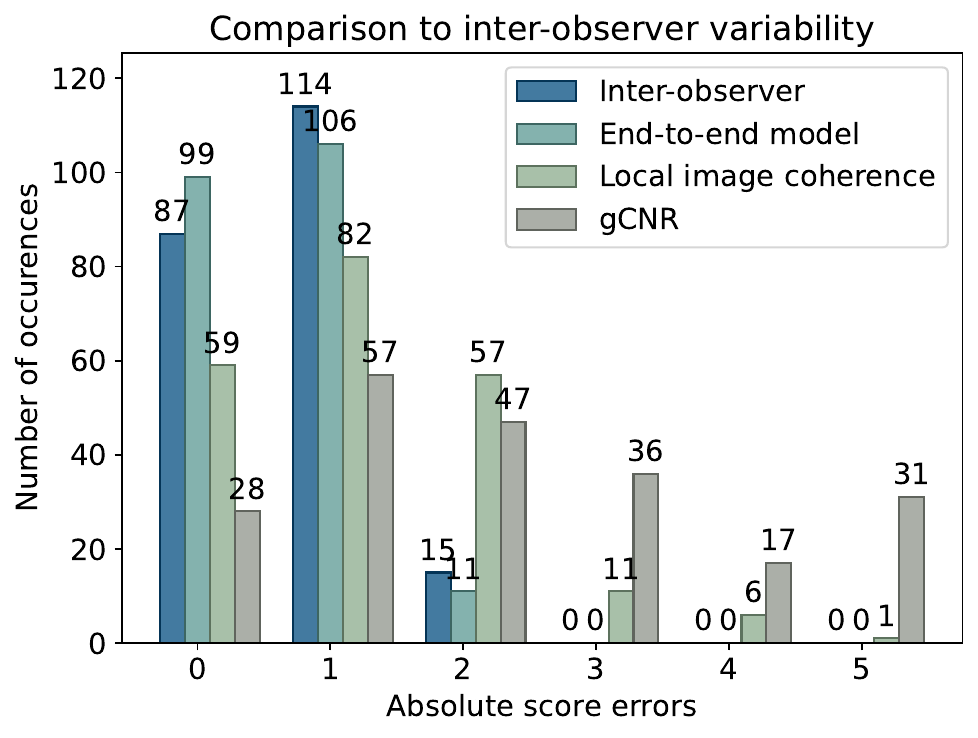}
  \caption{Bar plot comparing inter-observer variability to automatic methods. A method with lower variability will have the most occurrences with low score errors. Here we can observe that the variability of the end-to-end model is on par with the inter-observer variability, while the two other methods are not.
  }
  \label{fig: comp_inter_observer_variability_diffs}
\end{figure}

Fig. \ref{fig: comp_inter_observer_variability_diffs} shows the bar plot comparing the automatic methods to the inter-observer variability and Table \ref{table: comp_inter_observer_variability_diffs} lists the corresponding average metric values. Using the Wilcoxon signed-rank test \cite{c4091bd3-d888-3152-8886-c284bf66a93a} and a significance level of $p=0.05$, we find that the difference in Mean absolute error (MAE) between each of the methods is statistically significant. The difference between the inter-observer MAE and the MAE of each of the methods is also statistically significant, i.e. the inter-observer MAE is higher than the MAE of the end-to-end model and lower than the MAE of the other two models.

\begin{table}
\scriptsize
  \centering
  \caption{Comparison of automatic methods to inter-observer variability on the first round of annotations. For this data, there are three annotation values for each quality label. Each of the automatic methods is compared against each of the three annotators and the results are averaged. For the inter-observer variability, the labels are compared pairwise between each of the three annotators.}
  \begin{tabular}{m{100pt}|m{100pt}|m{100pt}|m{100pt}}
  \toprule
     Method & Spearman correlation & MAE & Accuracy \\
    \midrule
    End-to-end model & \textbf{0.69} & \textbf{0.59$\pm$0.59} & \textbf{46\%} \\
    Local image coherence & 0.58 & 1.19$\pm$1.01 & 27\% \\
    gCNR &0.24 & 2.23$\pm$1.58 & 13\%\\
    \midrule
    Inter-observer & 0.63 & 0.67$\pm$0.60 & 40\% \\
    \bottomrule
  \end{tabular}
      \label{table: comp_inter_observer_variability_diffs}
\end{table}

\subsection{Results of relation to variability in clinical measurements} \label{subsection: Results of relation to variability in clinical measurements}

\begin{figure}
     \centering
     \begin{subfigure}[b]{0.49\linewidth}
         \centering \includegraphics[trim={0cm 0cm 1.5cm 0cm}, clip, width=1\linewidth]{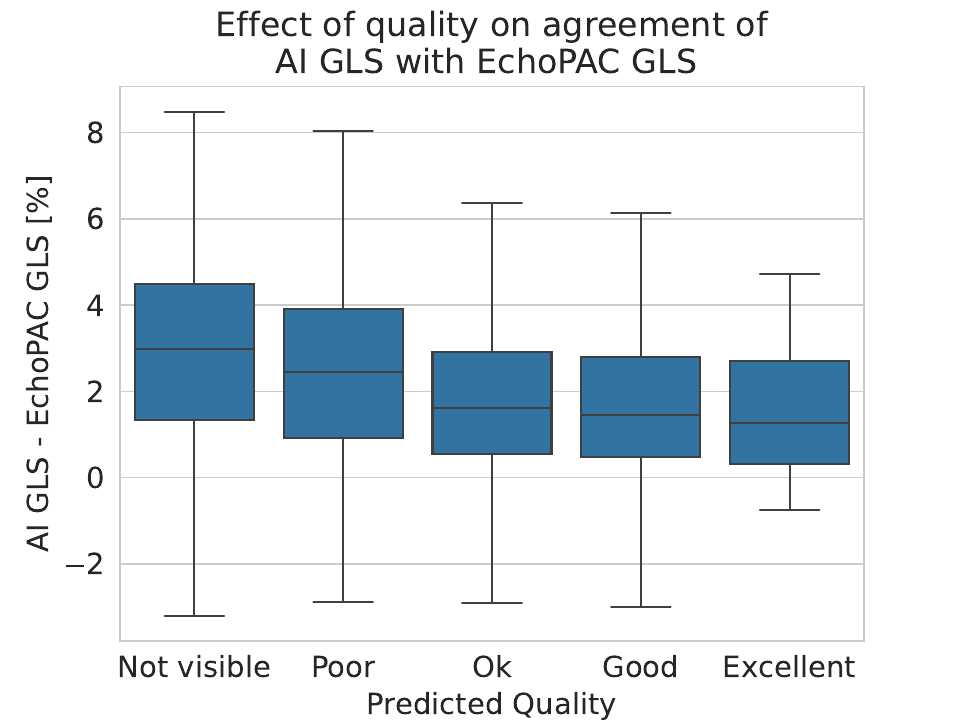}
         \caption{AI GLS \cite{ostvik2021myocardial} vs EchoPAC GLS (2D Strain tool)\cite{nyberg2023echocardiographic}. }
         \label{fig: PGLS_ref}
     \end{subfigure}
     \begin{subfigure}[b]{0.49\linewidth}
         \centering \includegraphics[trim={0cm 0cm 1.5cm 0cm}, clip, width=1\linewidth]{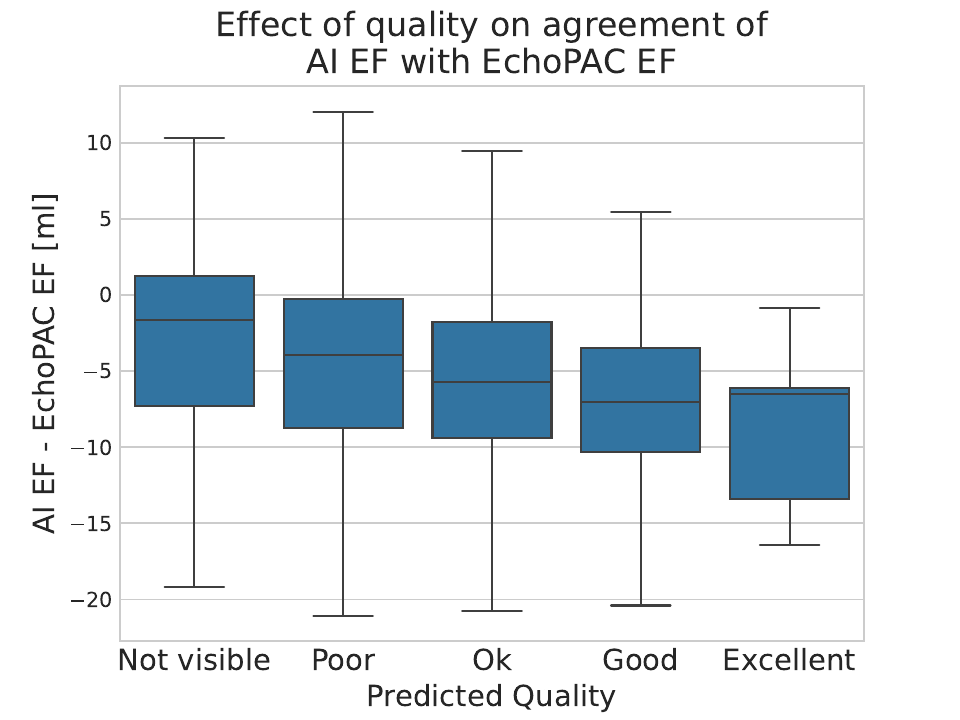}
         \caption{AI EF \cite{9037081} vs EchoPAC EF \cite{olaisen2023automatic}.}
         \label{fig: EF_ref}
     \end{subfigure}
   \caption{Box plots of the difference between clinical measurement values obtained automatically by AI \cite{ostvik2021myocardial, 9037081} and reference measurements obtained manually using GE HealthCare EchoPAC on the HUNT4 data \cite{nyberg2023echocardiographic, olaisen2023automatic} per image quality category, as predicted by the end-to-end model. The decrease in standard deviation with better image quality indicates a better agreement between the methods for higher image quality. Additionally, there is a noticeable change in bias between different quality categories. We believe this effect is partly caused by physiological differences correlated with image quality and is out of scope for this work.}
    \label{fig: measurements_quality}
\end{figure}

Fig. \ref{fig: measurements_quality} shows box plots visualizing the agreement between the measurements obtained automatically and with EchoPAC for each predicted quality category. The standard deviations in these plots represent how well the AI estimates agree with the manual references. The main finding is that the limits of agreement are narrower for higher qualities.

%% file: Sections/Discussion.tex
\section{Discussion}

\subsection{Challenges and considerations}

Assessing image quality based on human perception is inherently a subjective task, even when supported by clear definitions of image quality categories. This creates challenges for training and evaluation as there is no ground truth as the reference labels are a subjective estimation themselves. Therefore, it is not realistic to expect the automatic models to agree with reference labels as well as on tasks with well-defined ground truth labels. This, together with a rather fine scale of image quality categories, explains the low accuracies in Tables \ref{table: quality_metrics_results} and \ref{table: comp_inter_observer_variability_diffs}. Fig. \ref{fig: comp_inter_observer_variability_diffs} shows how the end-to-end model has on average less error than the annotators between each other. This indicates the model has learned to produce quality labels that strike a middle ground between the subjective assessments of the annotators. \newline

The end-to-end learning model overestimates low-quality regions and underestimates high-quality regions, as can be seen in Fig. \ref{fig: box_plots_end_to_end}. This means that the model can only explain a limited amount of variability in the image quality labels and is a result of minimizing the mean squared error (MSE) while dealing with subjective, and thus noisy reference values with fixed boundaries. We can eliminate this effect by fitting a linear model on the validation set that maps predicted image quality to image quality labels and applying it when doing inference on the test set. This increases MSE but gives more uniform performance over the image quality labels.\newline

One reason for the weak correlation between the annotations and the pixel-based methods is the rough selection of ROI and background region. Fig. \ref{fig: box_plots} shows how the average metrics of the classic pixel-based and coherence metrics increase for each quality label until the \textit{good} label, and then drop again for \textit{excellent}.  This is because on the one hand in these high-quality images, the blood speckles can be visible inside the LV lumen, which is used as background region, and on the other hand the myocardium tissue, which is used as ROI, is less blurred resulting in a smaller spread of pixels with high intensity. This can be seen for the anterolateral wall and apex in the rightmost column of Fig. \ref{fig: example_quality_cases}. One possibility to only select regions belonging to the tissue is to perform automatic pixel selection methods like Otsu thresholding \cite{otsu1979threshold} or percentile filters, but in our experiments this reduced the performance even further.

\subsection{Design choices}

The different methods in this study have a trade-off between accuracy and versatility. The default end-to-end network gives the best results but requires specific image quality labels for the task. Next, the coherence-based method is more generic and can potentially be applied more generally without the need for view-specific image quality annotations. Rindal et al. \cite{rindal2023very} showed that the GIC is not significantly different between apical views, but is higher for apical views than parasternal views. Thus, while a single image-to-image model can learn to predict coherence for different views, the mapping from coherence to image quality should be done for each group of views separately. Another advantage is that coherence can be used to give a global image quality metric without the need for a segmentation model. Finally, the pixel-based methods can be applied automatically in the most general way given a segmentation model to select ROI and background regions but also give the lowest accuracy. \newline

The ablation study of the end-to-end learning model showed that increasing the complexity of the model did not improve the performance. This is a result of the relatively small dataset size and the specific task of the model. For a more general model of image quality prediction with more varied input, e.g. one model for all cardiac views, a larger dataset and more complex model may be required.

\subsection{Clinical use}

Image quality estimation can be the first step towards a method for giving reliability estimates to clinical measurements and quality control of fully automatic methods. Fig. \ref{fig: measurements_quality} shows that the variability in clinical measurements goes down with higher predicted quality. However, image quality is only one source of variability, so a reliability model would also need to include view correctness and other factors that determine whether a given input is difficult to assess. \newline 

More direct use cases of the quality prediction model include the automatic selection of the best frame to perform a clinical measurement when multiple options are available, data cleansing in data mining, and automatic disapproval of segments for regional strain analysis. All the methods explored in this work are computationally efficient and can be run in real-time while scanning, and can thus be used as a guidance tool to enable clinicians to acquire images with better image quality.

\section{Real-time demo application}

\begin{figure}
\centering
  \centering
  \includegraphics[trim={0.5cm 2.5cm 5.25cm 3cm}, clip, width = 0.55\linewidth]{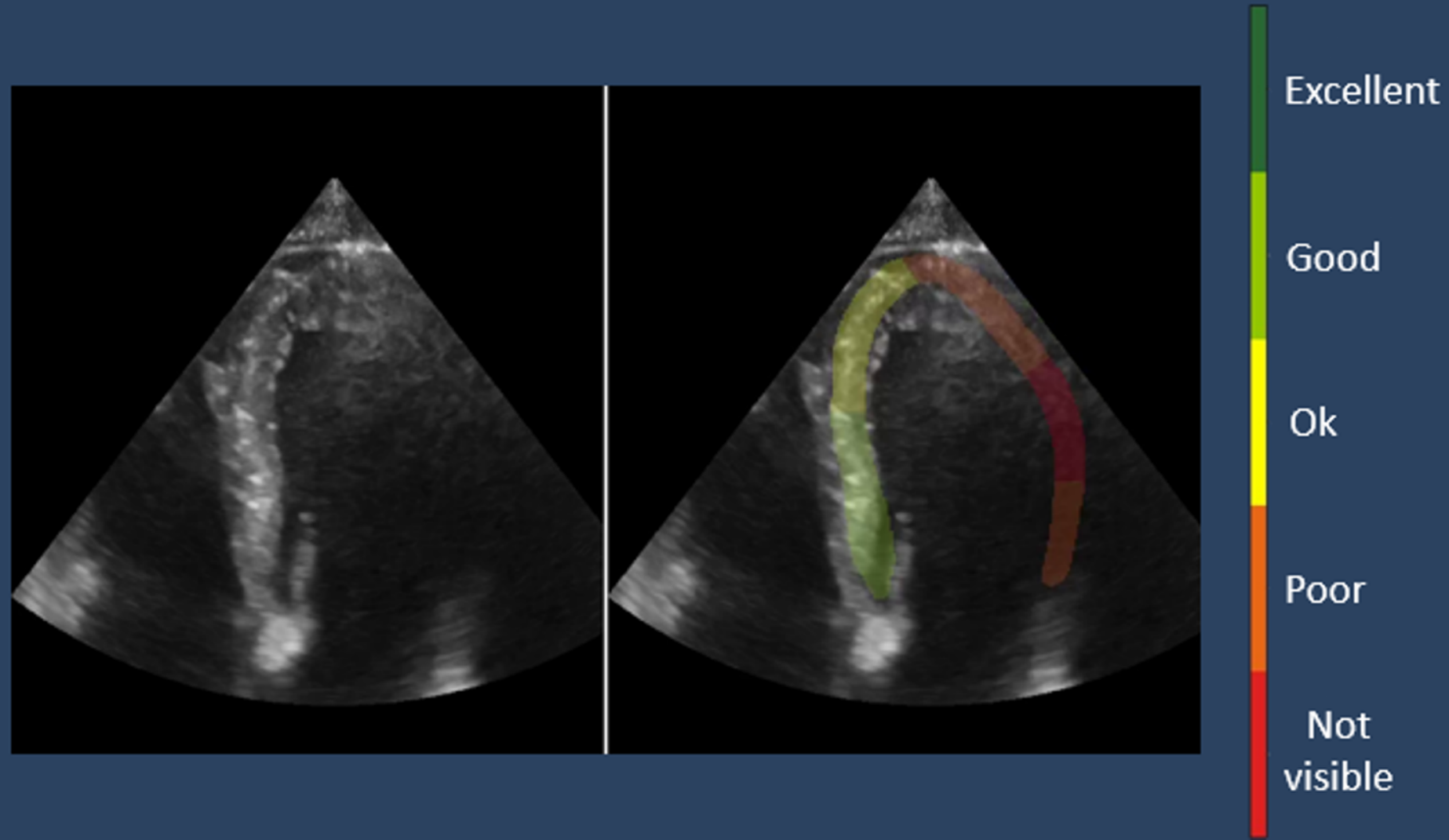}
  \caption{Screenshot of the real-time demo application. The left side shows the input B-Mode image. The right side shows the output of the segmentation color-coded by the output of the end-to-end quality network. The color codes are the same as in Fig. \ref{fig: example_quality_cases}.
  }
  \label{fig: demo_screenshot}
\end{figure}

To showcase the functionality of the end-to-end, real-time quality network, a real-time application was developed using the FAST framework \cite{smistad2015fast}. The demo is a split-screen application that shows the B-Mode input to the left and the segmentation regionally color-coded by the quality as predicted by the end-to-end network to the right. Fig. \ref{fig: demo_screenshot} provides a screenshot of the application in use. We provide a demo video \cite{VanDeVyver2024} illustrating the application in action while a clinician operates a GE Vivid E95 scanner. The video can be accessed at \url{https://doi.org/10.6084/m9.figshare.26413984}.

%% file: Sections/Conclusion.tex
\section{Conclusion}

In this work, we developed and compared different deep-learning methods for regional image quality estimation in cardiac ultrasound. We show that classic pixel-based methods, such as (g)CNR, together with automatic image segmentation, give low agreement with the quality assessment of cardiologists. We developed a U-Net model to predict the coherence factor for each pixel in the ultrasound image and showed that the resulting coherence image can be used to assess the image quality in a pixel-based way with better performance than the classic measures. The best results, below inter-observer variability, are obtained by using an end-to-end deep-learning model. Finally, we show higher predicted quality is associated with lower limits of agreement between fully automatic and manual methods for the clinical measurements EF and GLS. \newline

%% file: Sections/appendices.tex
\appendix

\section{Extraction of cardiac regions of interest} 
\label{appendix: Extraction of cardiac regions of interest}

We use the nnU-Net \cite{isensee2021nnu, nnUNetV2} architecture to segment the cardiac images. The nnU-Net is used out of the box using the default configuration but without the final ensemble step. Instead, we train and validate on a single predefined 80\% train, 10\% validation, and 10\% test split from the HUNT4 segmentation annotation dataset. Table \ref{table: characteristics nnunet} summarizes the characteristics of the nnU-Net architecture. This model is described in more detail and compared to other segmentation models in our previous work \cite{10458930}. We use two nnU-Nets, one for apical two- (A2C) and four-chamber (A4C) views, and one for apical long axis (ALAX) views. The nnU-Net for A2C and A4C views segments the left ventricle (LV), left atrium (LA) and myocardium (MYO). The nnU-Net for ALAX views additionally segments the aorta (AO).
\newline

\begin{table}
\scriptsize
  \centering\caption{Characteristics of the nnU-Net \cite{isensee2021nnu, nnUNetV2} used. The "number of channels" row indicates the number of channels at the first, bottom, and last convolution of the U-Net respectively. The number of output channels reflects the number of structures the model segments. For A2C/A4C views, the nnU-Net segments the LV, MYO, and LA. For ALAX views, the nnU-Net additionally segments the AO.}
  \begin{tabular}{m{100pt}|m{250pt}}
    \toprule
    Input size & $256 \times 256$ \\
    Number of channels & 32 $\downarrow$ 512 $\uparrow$ 32 \\ 
    Number of output channels & 3 for A2C/A4C views, 4 for ALAX views \\ 
    Lowest resolution & $4 \times 4$ \\ 
    Upsampling scheme & Deconvolutions \\ 
    Normalization scheme & InstanceNorm \\ 
    Batch Size & 49 \\
    Optimizer & Adam \\
    Initial learning Rate & 1e-2 \\ 
    Scheduler & Polynomial \\ 
    Loss function & Dice \& cross-entropy\\ 
    Inter-layer activation & Leaky Relu \\ 
    Final layer Activation & Softmax \\ 
    Epochs & 500 \\
    Augmentations & Rotations, scaling, Gaussian noise, Gaussian blur, brightness, contrast, simulation of low resolution, gamma correction and mirroring. For more details, see \cite{nnUNetV2}. \\
        \bottomrule
  \end{tabular}
      \label{table: characteristics nnunet}
\end{table}

\begin{figure}
     \centering
     \begin{subfigure}[b]{0.225\linewidth}
         \centering
\includegraphics[width=1\linewidth]{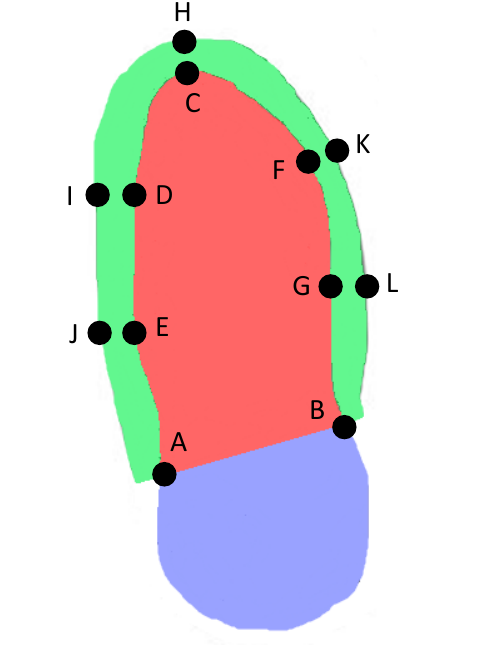}
         \caption{Point extraction}
         \label{fig: region_division_a}
     \end{subfigure}
     \begin{subfigure}[b]{0.225\linewidth}
         \centering \includegraphics[width=1\linewidth]{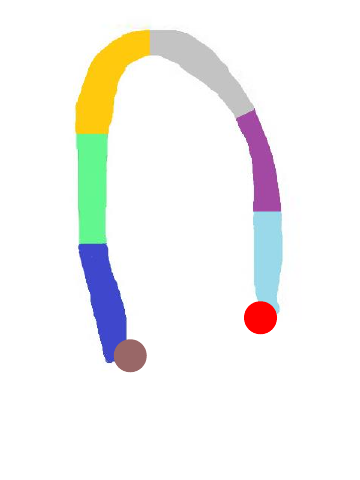}
         \caption{MYO divided into \newline \hspace*{0.45cm} regions}
         \label{fig: region_division_b}
     \end{subfigure}
     \hspace{1cm}
          \begin{subfigure}[b]{0.225\linewidth}
         \centering
\includegraphics[trim={4cm 0cm 4cm 0cm},clip,width=1\linewidth]{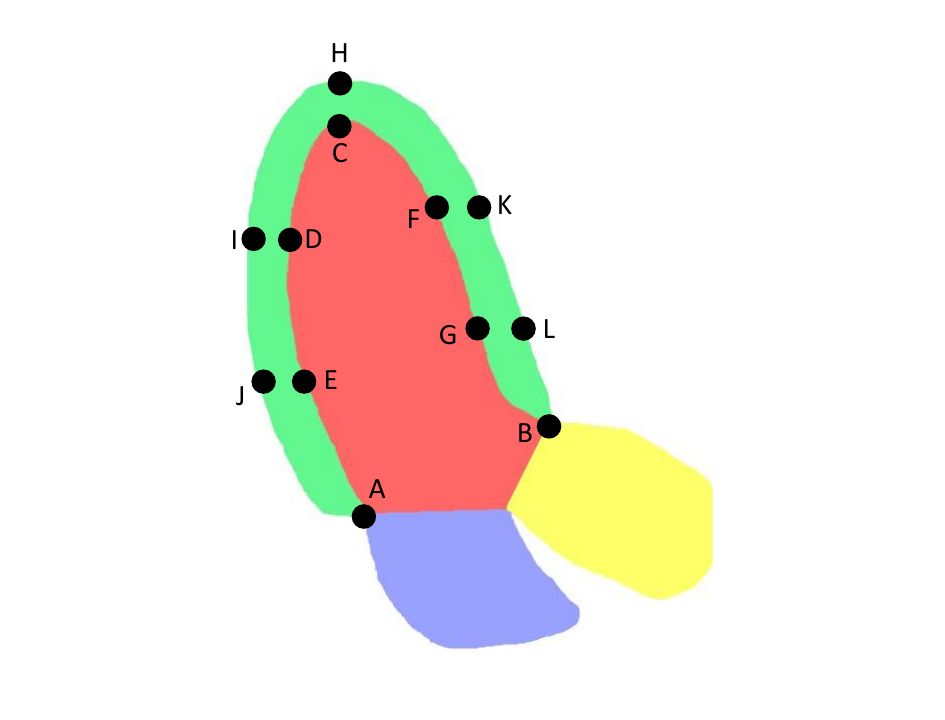}
         \caption{Point extraction}
         \label{fig: region_division_alax_a}
     \end{subfigure}
     \begin{subfigure}[b]{0.225\linewidth}
         \centering \includegraphics[trim={4cm 0cm 4cm 0cm},clip,width=1\linewidth]{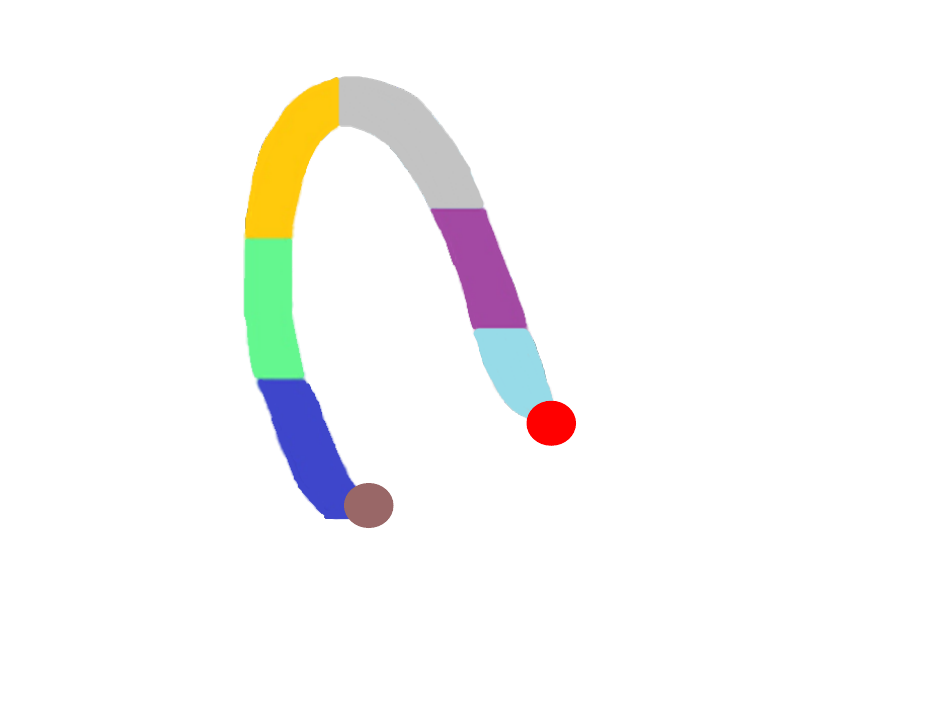}
         \caption{MYO divided into \newline \hspace*{0.45cm} regions}
         \label{fig: region_division_alax_b}
     \end{subfigure}
        \caption{Division of MYO into regions. A and B represent the MYO base points. C represents the LV apex. D, E, F and G are obtained by dividing the inner LV border into equal parts. H, I, J, K and L are obtained by finding the closest corresponding point on the outer MYO border. }
        \label{fig: region_division}
    \hspace{1cm}
\end{figure}

The segmentation of the MYO is divided into eight regions using the following algorithm:
\begin{enumerate}
  \item Extract the annulus points. For A2C/A4C views, these are the points where the MYO meets the LA. For ALAX views, these are the points where the MYO meets the LA and AO. Points A and B are the annulus points in Fig.~\ref{fig: region_division}.
  \item Extract the apex of the LV, defined as the furthest points from the base points within the lv lumen. This is point C in Fig.~\ref{fig: region_division}.
  \item Divide both the left and right part of the endocardium border, defined as the border between the LV and MYO regions, into three parts with equal length. This gives points D, E, F and G in Fig.~\ref{fig: region_division}.
  \item Find the closest points on the outer MYO border for points C, D, E, F and G. These are points H, I, J, K and L in Fig.~\ref{fig: region_division}.
  \item Fill in the regions by connecting the points via the contour, resulting in the MYO divided into six regions. 
  \item Draw circles\footnote{Due to the unequal pixel spacing in depth and width, the annulus regions become ovals in the 256x256 segmentation maps. When plotting the images with equal spacing in width and depth, these regions become circles again.} with a radius of 2  millimeters around the annulus points, points A and B in Fig.~\ref{fig: region_division}. 
  We use these additional two regions to asses the local image quality of the annulus points in the image. The result are the eight regions, as in Fig.~\ref{fig: region_division}.
  \item Remove any parts of regions that fall outside of the sector. The apical top regions in~Fig.\ref{fig: otsu_thresholding_example_density_a}, i.e., the yellow and white masks, are examples of this. If more than 50\% of all pixels inside the region fall outside the sector, we exclude the region from analysis.
\end{enumerate}
The goal is to automatically quantify the image quality in each of these eight regions. The LV lumen is used as background region.

\section{Ablation study of the end-to-end learning model}
\label{appendix: Ablation study end-to-end learning model}

\begin{figure}
     \centering
     \begin{subfigure}[b]{0.3\linewidth}
         \centering \includegraphics[trim={2cm 1.4cm 1cm 1cm}, clip, width=1\linewidth]{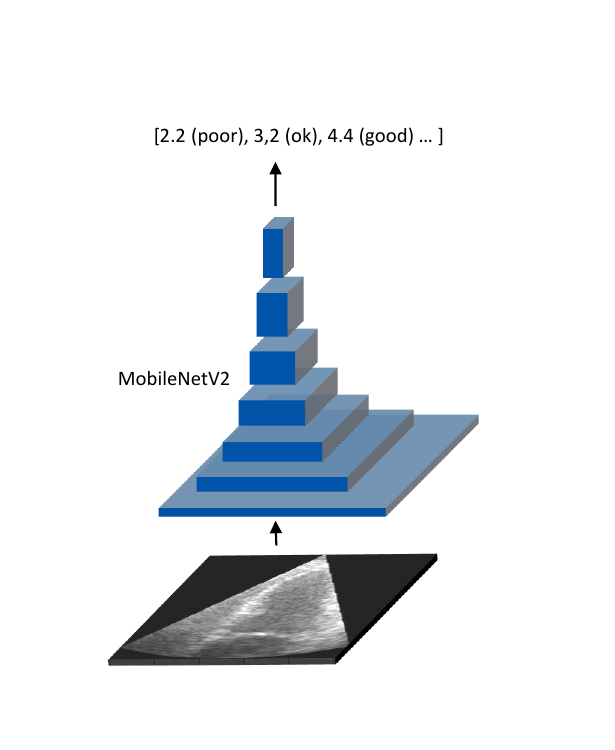}
         \caption{\raggedright{Default end-to-end \newline \makebox[0.55cm][l]{~}model}}
         \label{fig: model_variants_a}
     \end{subfigure}
          \hspace*{-1cm}
     \begin{subfigure}[b]{0.3\linewidth}
         \centering \includegraphics[trim={2cm 1.4cm 1cm 1cm} ,clip,width=1\linewidth]{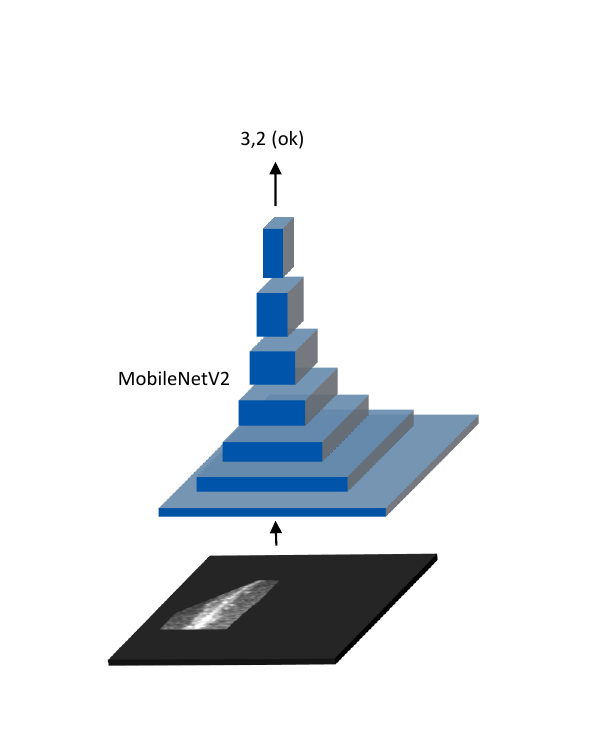}
         \caption{\raggedright{End-to-end hard \newline \makebox[0.55cm][l]{~}attention model}}
         \label{fig: model_variants_b}
     \end{subfigure}
          \hspace*{-1cm}
     \begin{subfigure}[b]{0.38\linewidth}
        \centering \includegraphics[trim={0.5cm 0.7cm 0.3cm 0cm}, clip,width=1\linewidth]{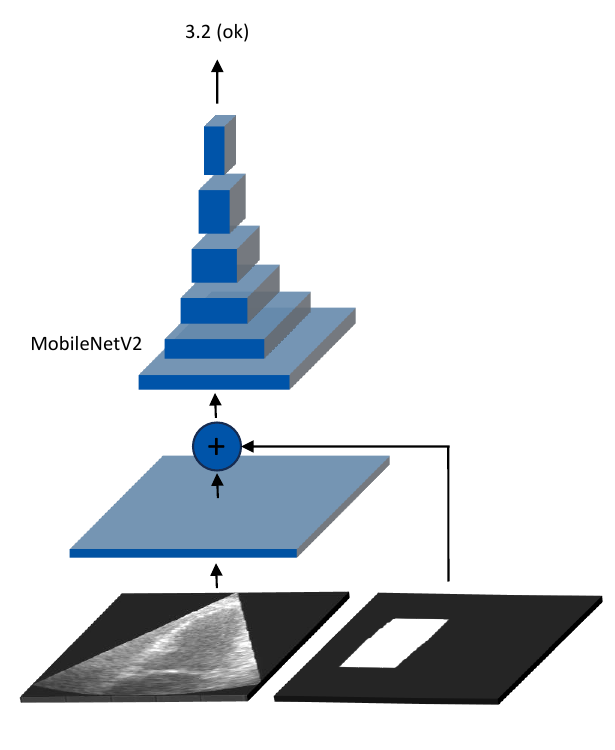}
         \caption{\raggedright{End-to-end soft attention model \newline}}
         \label{fig: model_variants_c}
     \end{subfigure}
        \caption{Different variants of the end-to-end learning model. The default model outputs a list of all quality scores at once, while the other two variants output only a single quality score. The end-to-end soft attention model corresponds to version 'c' in Eppel \cite{eppel2018classifying}.}
        \label{fig: model_variants}
    \hspace{1cm}
\end{figure}

In the ablation study, we evaluated the impact of modifying the architecture of the end-to-end learning model. Three different network architectures were tested: Cardiac View Classification (CVC) network \cite{ostvik2019real}, MobileNetV2 \cite{sandler2018mobilenetv2}, and EfficientNet \cite{tan2019efficientnet}. 
We tested approaching the problem both as a classification and regression task, with only the final dense layer and loss function being changed accordingly. Additionally, three basic network attention variations were tested using the automatic segmentation output. The default model did not use any attention and predicts each label directly, as in~Fig.\ref{fig: model_variants_a}. For the other two versions, the region masks extracted from the segmentation were dilated with a square dilation filter of size 50x50 pixels and used as an additional input to the networks which then predict the label of one region at a time. The dilation filter reveals the direct vicinity around each region so the boundary between tissue and background becomes visible. The first variant used this dilated mask as hard attention by blacking out the other parts of the image, as shown in Fig.~\ref{fig: model_variants_b}. For the second variant, the masks were used as soft attention as input to a side branch of the network, as proposed by Eppel \cite{eppel2018classifying}. For this version, we created an attention map and added it element-wise to the output of the first layer, corresponding to version 'c' in Eppel \cite{eppel2018classifying}. Fig.~\ref{fig: model_variants_c} shows this configuration. \newline

The ablation study consists of two parts. Both parts used the training configuration listed in Table \ref{table: train_params_end-to-end} and data from the regional image quality dataset. In the first part, we examined the effect of changing the convolutional backbone and the effect of framing the problem as a classification or regression task. Table \ref{table: ablation_study_1} compares the predictions with the annotations on the test set for the different configurations using the default end-to-end model. In the second part of the ablation study, the backbone was fixed to MobileNetV2 \cite{sandler2018mobilenetv2} and the problem was framed as a regression task. Next, the different variants shown in Fig. \ref{fig: model_variants} were compared. Table \ref{table: ablation_study_2} summarizes the results on the test set. \textit{The default end-to-end model with no attention, the MobileNetV2 \cite{sandler2018mobilenetv2} architecture, and the problem framed as a regression task gave the best results and were thus used for this paper}.

\begin{table}
\scriptsize
  \centering
  \caption{Training configuration of the end-to-end learning model}
  \begin{tabular}{m{100pt}|m{250pt}}
    \toprule
    Input size & 256x256 \\
    Batch size & 16 \\
    Optimizer & Adam \\
    Initial learning rate& 1e-4 \\
    Scheduler & None \\
    Loss function regression & Mean squared error (MSE) \\
    Loss function classification & Squared earth mover’s distance \cite{hou2016squared} \\
    Epochs & 500 \\
    Augmentations & Rotations ($-30^{\circ} \leq \text{angle} \leq  30^{\circ}$), horizontal mirroring (also mirror labels), gamma correction ($0.9\leq \gamma \leq 1.1$) and scaling ($0.85 \leq \text{magnification} \leq 1.15$). Each augmentation is applied individually with $0.5$ probability.\\
    \bottomrule
  \end{tabular}
      \label{table: train_params_end-to-end}
\end{table}

\begin{table}
\scriptsize
  \centering
  \caption{Results on the test set for the first part of the end-to-end learning ablation study. The Cardiac View Classification (CVC) \cite{ostvik2019real} network is trained from scratch. The other networks are pretrained on ImageNet \cite{deng2009imagenet}. Each method is compared against annotations from the test set of the regional image quality dataset.}
  \begin{tabular}{m{60pt}|m{75pt}|m{75pt}|m{100pt}|m{75pt}}
    \toprule
    Task & Model  & Spearman correlation & Mean absolute error (MAE) & Number of parameters in millions \\
    \midrule
    \multirow{5}{1.4cm}{Regression} & CVC \cite{ostvik2019real} & 0.62 & 0.60$\pm$ 0.45 & 9.6  \\
    & MobileNetV2 \cite{sandler2018mobilenetv2} & \textbf{0.65} & \textbf{0.59$\pm$0.45} & 2.2\\
    & EfficientNet-B0 \cite{tan2019efficientnet} & 0.63 & 0.61$\pm$ 0.45 & 4.0  \\
    & EfficientNet-B4 \cite{tan2019efficientnet} & 0.63 & 0.60$\pm$0.44 & 18  \\
    & EfficientNet-B7 \cite{tan2019efficientnet} & 0.63 & 0.59$\pm$0.46 & 64 \\
    \midrule
    \multirow{5}{1.4cm}{Classification} & CVC \cite{ostvik2019real} & 0.38 & 0.91$\pm$ 0.81 & 9.7  \\
    & MobileNetV2 \cite{sandler2018mobilenetv2} & 0.48 & 0.80$\pm$0.71 & 2.3\\
    & EfficientNet-B0 \cite{tan2019efficientnet} & 0.46 & 0.80$\pm$ 0.76 & 4.1 \\
    & EfficientNet-B4 \cite{tan2019efficientnet} & 0.46 & 0.83$\pm$0.73 & 18  \\
    & EfficientNet-B7 \cite{tan2019efficientnet} & 0.37 & 0.91$\pm$0.79 & 64 \\
    \bottomrule
  \end{tabular}
      \label{table: ablation_study_1}
\end{table}

\begin{table}
\scriptsize
  \centering
  \caption{Results on the test set for the second part of the end-to-end learning ablation study using MobileNetV2. }
  \begin{tabular}{m{100pt}|m{75pt}|m{100pt}}
  \toprule
    Model variant & Spearman correlation & Mean absolute error (MAE) \\
    \midrule
    Default (no attention) & \textbf{0.65} & \textbf{0.59$\pm$0.45} \\
    Variant 1 (hard attention) & 0.47 & 0.68$\pm$0.60\\
    Variant 2 (soft attention) & 0.53 & 0.67$\pm$0.57\\
    \bottomrule
  \end{tabular}
      \label{table: ablation_study_2}
\end{table}